\documentclass{emulateapj}
\usepackage{apjfonts}
\usepackage{amsmath}
\usepackage{lscape}
\usepackage{mathrsfs}
\usepackage{epsfig}

\journalinfo{The Astrophysical Journal in press}
\slugcomment{Received 2008 December 11; accepted 2009 April 15}
\shorttitle{CONSTRAINING SPINS OF SMBHS FROM TEV VARIABILITY: RELATIVISTIC CALCULATIONS}
\shortauthors{LI ET AL.}

\def\tautev{\tau_{_{\rm TeV}}}
\def\ttev{\Theta_{\rm TeV}}
\def\tobs{\Theta_{\rm obs}}
\def\rtev{R_{\rm TeV}}
\def\rd{{\rm d}}

\begin{document}

\title{Constraining spins of supermassive black holes from TeV variability. II. fully general relativistic 
calculations}

\author{Yan-Rong Li\altaffilmark{1}, Ye-Fei Yuan\altaffilmark{2},
 Jian-Min Wang\altaffilmark{1,4}, Jian-Cheng Wang\altaffilmark{3},
 \and Shu Zhang\altaffilmark{1}}
\affil{1 Key Laboratory for Particle Astrophysics, Institute of High Energy Physics, CAS, 
19B Yuquan Road, Beijing 100049, China\\
2 Key Laboratory for Research in Galaxies and Cosmology, University of Sciences and 
Technology of China, CAS, Hefei, Anhui 230026, China\\
3 Yunnan Observatory, CAS, Kunming 650011, China \\
4 Theoretical Physics Center for Science Facilities (TPCSF), CAS}

\begin{abstract}
The fast variability of energetic TeV photons from the center of M87 has been detected, offering
a new clue to estimate spins of supermassive black holes (SMBHs). We extend the study of Wang et al. 
(2008) by including all of the general relativistic effects. We numerically solve the full set of 
relativistic hydrodynamical equations of the radiatively inefficient accretion flows (RIAFs)
and then obtain the radiation fields around the black hole. The optical depth of the radiation 
fields to TeV photons due to pair productions is calculated in the Kerr metric. We find that 
the optical depth strongly depends on (1) accretion rates as $\tautev\propto \dot{M}^{2.5-5.0}$;  
(2) black hole spins; and (3) location of the TeV source. Jointly considering the optical 
depth and the spectral energy distribution radiated from the RIAFs, the strong degeneration of the spin
with the other free parameters in the RIAF model can be largely relaxed. We apply the present model to 
M87, wherein the RIAFs are expected to be at work, and find that the minimum specific angular 
momentum of the hole is $a\sim0.8$. The present methodology is applicable to M87-like sources with 
future detection of TeV emissions to constrain the spins of SMBHs. 
\end{abstract}

\section{Introduction}
Astrophysical black holes can be  simply depicted by two parameters: mass and angular momentum. Masses 
of supermassive black holes (SMBHs) are relatively easier to estimate since their gravitational influences 
spread over the large-scale space approached by Newtonian mechanics. SMBH masses in active galactic nuclei 
or galactic centers can be measured by several different methods through stellar or gas dynamics, or 
reverberation mapping of emission lines (Kaspi et al. 2000; Kormendy \& Gebhardt 2001; Ho 2008). 
However, spins are more elusive to estimate because their general relativistic (GR) effects only appear 
significantly in the innermost region close to the black hole, typically  within $\sim$20 gravitational 
radii and eventually disappear outward then. Spatially resolving the region around $\sim$10 
or a few gravitational radii is in principle plausible through radio observations for the 
Galactic center (Shen et al. 2005; Doeleman et al. 2008), but it is still a limit challenge to current 
radio astronomy for estimation of the spins of the extragalactic SMBHs.

Hitherto, only a few Seyfert galaxies show relativistically broadened, highly redshifted iron K$\alpha$ 
emission lines, which are interpreted most plausibly by the collective effects of the Doppler motion of 
the fluids from which the intrinsic narrow emission lines originate, the gravitational redshift, and the 
gravitational lensing around a rapidly rotating black hole ({see, e.g., Miller 2007 and references therein}). 
The most convincing evidence for effects of spins is found in the MCG-6-30-15 from XMM-{\em Newton}
observations (Fabian et al. 2002). On the other hand, cosmic X-ray background radiation suggests that most 
SMBHs are fast rotating (Elvis et al. 2002) and the Soltan's argument applied to large samples of quasars 
and galaxies indicates a similar conclusion (Fabian \& Iwasawa 1999; Yu \& Tremaine 2002; Wang et al. 2006).

TeV photons suffering from attenuation by pair productions are expected to explore the radiation 
fields in the vicinity of SMBHs and thus constrain their spins (Wang et al. 2008, hereafter Paper I). 
Rapid variability of TeV emission (at a timescale of 2 days) 
has been discovered by the HESS (High Energy Stereoscopic System) collaboration in the famous radio
galaxy M87 (Aharonian et al. 2006). Interestingly, the TeV emission does {\em not} originate from a 
relativistic jet (Aharonian et al. 2006), in contrast to cases of blazars 
(Blandford \& Levinson 1995; Levinson 2006). A novel mechanism for TeV emission around 
the horizon of a spinning black hole by a magnetospheric pulsar-like process was originally proposed by 
Levinson (2000), and had been applied to M87 (Neronov \& Aharonian 2007, but see Rieger \& Aharonian 2008). 
In such a context, the TeV photons are able to serve as a probe of spins which determine the density 
of the radiation fields.  Wang et al. (2008) found that the SMBH in M87 should have
specific angular momentum $a>0.65$ in order to allow the TeV photons to escape from the innermost region
of the radiation fields from the accretion disk. However, their calculations are based on the self-similar 
solutions of ADAFs in Newtonian approximation and the GR effects on the disk 
structures and the propagation of photons are not included.

The main goal of this paper is to extend the study of Paper I by including all the GR effects. 
In Section 2, we introduce the GR RIAF model. Section 3 gives the description of relativistic optics and
the detailed formulae used for calculation of optical depth to TeV photons. The results are  presented
in Section 4 with exhaustive investigation of parameter dependence of the optical depth. We then apply the results  
to M87 in Section 5. Discussions and a summary are given in Section 6 and 7, respectively. Mathematical 
preliminaries are presented in the Appendix.

\section{General Relativistic RIAF model}
We follow the work of Manmoto (2000) to construct the fully relativistic hydrodynamical equations. GR 
notations are given in the Appendix. Cylindrical coordinates ($t, R, \phi, z$) are used to describe accretion 
flows by expanding the Boyer$-$Lindquist coordinates around the equatorial plane up to $(z/R)^0$ terms. All the 
basic equations can be derived from the conservation laws of mass, momentum, and energy under the four- 
dimensional space. For a clarification, we list the necessary equations as follows. 

The continuity equation reads
\begin{equation}
 \dot M=-2\pi\Delta^{1/2}\Sigma_0\gamma_rV,
\end{equation}
where $V$ is the radial velocity of the flows with respect to the corotating reference frame (CRF, see the Appendix) 
with $\gamma_r$ as its Lorentz factor (e.g., Abramowicz et al. 1996; Gammie \& Popham 1998; Yuan et al. 2009), 
$\dot M$ is the mass accretion rate, and $\Sigma_0$ is the surface 
density of accretion flows.  The conservation of momentum gives two equations. The radial component is 
\begin{equation}
\gamma_r^2V\frac{\rd V}{\rd R}=-\frac{1}{\mu\Sigma_0}\frac{\rd W}{\rd R}-
                                \frac{\gamma_\phi^2 AM_{\bullet}}{R^4\Delta}
			        \frac{(\Omega-\Omega_{\rm K}^{+})(\Omega-\Omega_{\rm K}^{-})}
			         {\Omega_{\rm K}^{+}\Omega_{\rm K}^{-}},
\end{equation}
where $\gamma_{\phi}$ is the Lorentz factor of the azimuthal velocity of the CRF with respect to the locally
 nonrotating frame (LNRF), the height-integrated total pressure $W$ consists of gas pressure and 
magnetic pressure, and the relativistic enthalpy $\mu$ is written as 
\begin{equation}
 \mu=1+\frac{W}{\Sigma_0}\left[\left(
a_i+\frac{1}{\beta_{\rm d}}
\right)\frac{W_i}{W}
+
\left(a_e+\frac{1}{\beta_{\rm d}}
\right)\frac{W_e}{W}\right].
\end{equation}
Here we express the functions $a_i$ and $a_e$ as
\begin{equation}
a_i=\frac{1}{\gamma_i-1}+\frac{2(1-\beta_{\rm d})}{\beta_{\rm d}},
\end{equation}
\begin{equation}
a_e=\frac{1}{\gamma_e-1}+\frac{2(1-\beta_{\rm d})}{\beta_{\rm d}},
\end{equation}
with the adiabatic indices of gas $\gamma_i$ and $\gamma_e$ as
\begin{equation}
\gamma_i=1+\theta_i\left[\frac{3K_3(1/\theta_i)+K_1(1/\theta_i)}{4K_2(1/\theta_i)}-1\right]^{-1},
\end{equation}
\begin{equation}
\gamma_e=1+\theta_e\left[\frac{3K_3(1/\theta_e)+K_1(1/\theta_e)}{4K_2(1/\theta_e)}-1\right]^{-1},
\end{equation}
where the global parameter $\beta_{\rm d}$ is the ratio of the gas pressure to the total pressure, 
$\theta_i=kT_i/m_pc^2$ and $\theta_e=kT_e/m_ec^2$ are the dimensionless temperatures of ions ($T_i$)  
and electrons ($T_e$), respectively, $k$ is the Boltzmann constant, $m_p$ and $m_e$ are proton's and
electron's mass, and $K_1$, $K_2$, and $K_3$ are the modified Bessel functions. 
The azimuthal component of the conservation of momentum is
\begin{equation}
 \dot M(\ell -\ell_{\rm in})=2\pi RW_\phi^R,
\end{equation}
where $\ell$ is the specific angular momentum of accretion flows, $\ell_{\rm in}$ is the 
swallowed by the central black hole, and
\begin{equation}
 W_\phi^R=\alpha_{\rm d}\frac{A^{3/2}\Delta^{1/2}\gamma_\phi^3}{R^6}W,
\end{equation}
where $\alpha_{\rm d}$ is the viscosity parameter. The conservation of energy for ions and electrons gives
\begin{equation}
 -\dot MT_{i}\frac{\rd s_{i}}{\rd R}=-2\pi\alpha_{\rm d}(1-\delta)W\frac{\gamma_\phi^4A^2}{R^6}
                                     \frac{\rd \Omega}{\rd R}
                                     -2\pi R\Lambda_{ie},
\end{equation}
\begin{equation}
  -\dot MT_{e}\frac{\rd s_{e}}{\rd R}=-2\pi\alpha_{\rm d}\delta W\frac{\gamma_\phi^4A^2}{R^6}\frac{\rd\Omega}{\rd R}
                                      +2\pi R(\Lambda_{ie}-F^-),
\end{equation}
where $s_i$ ($s_e$) is the specific entropy of ions (electrons), $\Lambda_{ie}$ is the energy transfer 
rate from the ions to the electrons per unit surface area, and $F^-$ is the radiative cooling per unit surface 
area. $\delta$ is the fraction of the viscous dissipation which heats the electrons. 
To complete the set of equations, thermal dynamical relations among the entropy, the 
height-integrated pressure and the surface density for ions and electrons are given by
\begin{equation}
T_{i}\rd s_{i}=\frac{W_{i}}{\Sigma_{i}}\frac{1}{\Gamma_{i}-1}
               \left[\rd\ln W_{i}-\Gamma_{i}\rd\ln\Sigma_{0}+(\Gamma_{i}-1)\rd\ln R\right],
\end{equation}
\begin{equation}
T_{e}\rd s_{e}=\frac{W_{e}}{\Sigma_{e}}\frac{1}{\Gamma_{e}-1}
            \left[\rd\ln W_{e}-\Gamma_{e}\rd\ln\Sigma_{0}+(\Gamma_{e}-1)\rd\ln R\right],
\end{equation}
where the effective adiabatic indices are
\begin{equation}
\Gamma_{i}=1+\left[a_{i}\left(1+\frac{\rd\ln a_{i}}{\rd\ln T_{i}}\right)\right]^{-1},
\end{equation}
\begin{equation}
\Gamma_{e}=1+\left[a_{e}\left(1+\frac{\rd\ln a_{e}}{\rd\ln T_{e}}\right)\right]^{-1}.
\end{equation}
The vertical scale height $H$ of the accretion flows is taken as (see also Abramowicz et al. 1997)
\begin{equation}
H^2=\frac{\mu W}{\Sigma_{0}}\frac{r^4}{\ell^2-a^2(e^2-\mu^2)},
\end{equation}
where $e$ is the specific energy.
The simplification $\rd\ln H/\rd\ln R=1$ (Manmoto 2000) is used in deriving the energy equations.

We take into account three processes of the radiative cooling, i.e.,  the synchrotron radiation,  
the bremsstrahlung, and the multi-Comptonization of soft photons. The general description of cooling 
processes and relevant formulae have been presented by Narayan \& Yi (1995) and Manmoto (2000) in a 
more handy way. In our calculations we completely make use of the program of the Comptonization 
given by Coppi \& Blandford (1990). 

The outer boundary conditions of the GR RIAF model are imposed at a radius
$r_{\rm out}=R_{\rm out}/R_{\rm g}=2\times10^4$:
\begin{equation}
\Omega=0.8\Omega_{\rm K};~~~T_{i}=T_{e}=0.1T_{\rm vir},
\end{equation}
where $R_{\rm g}=GM_{\bullet}/c^2$ and $T_{\rm vir}$ is the virial temperature defined as
\begin{equation}
T_{\rm vir}=(\gamma-1)\frac{GM_{\bullet}m_p}{kR}\approx 1.1(\gamma-1)\times 10^9r_4^{-1}~{\rm K},
\end{equation}
with the adiabatic index $\gamma=5/3$ and $r_4=r/10^4$.  Electrons and ions have the same temperature at
the outer boundary. As shown by Manmoto et al. (1997), the outer 
boundary conditions have little influence on the structures of the GR RIAFs. We confirmed this effect 
and thus fix the outer boundary conditions as Equation (17). 

The choice of the eigenvalue $\ell_{\rm in}$ should satisfy the condition that the flows pass 
through the sonic point smoothly. The inner edge of accretion flows is elusive since it depends on the
accretion rate, and is nonaxisymmetric and time-variable (see, e.g., Krolik \& Hawley 2002). 
We use the horizon of the black hole as the inner edge for a conserved influence of the spins
on the optical depth of TeV photons. 
We show this effect of inner edge on uncertainties of spins in the following sections in details. 

\section{General relativistic optics}

\subsection{Ray tracing method}
We assume that the accretion flows are axisymmetric, so the radiation fields have the same symmetry, and
 only the motions on the $(R-\theta)$ plane are required in our calculations.
The general trajectories of photons in Kerr metric are described by three constants of motion 
(Bardeen et al. 1972). In terms of the photon's four-momentum, the conserved quantities are $E=-p_t$, 
the total energy at infinity; $L=p_\phi$, the component of angular momentum parallel to symmetry axis and 
$q=p_\theta^2+\cos^2\theta a^2p_t^2+p_\phi^2\cot^2\theta$. By defining $\lambda=p_\phi/E$ and ${\cal Q}=q/E^2$, 
the four-momentum of photons can be rewritten as
\begin{equation}
P_\mu=(p_t,p_r,p_\theta,p_\phi)=\left[-1,\pm\Delta^{-1}\sqrt{{\cal R}(r)},\pm\sqrt{\Theta(\theta)}, \lambda\right]E,
\end{equation}
where
\begin{equation}
{\cal R}(r)=r^4+(a^2-\lambda^2-{\cal Q})r^2+2[{\cal Q}+(\lambda-a)^2]r-a^2{\cal Q},
\end{equation}
\begin{equation}
\varTheta(\theta)={\cal Q}+a^2\cos^2\theta-\lambda^2\cot^2\theta,
\end{equation}
where $r=R/R_{\rm g}$.

Basically, the trajectory of a photon on the ($R-\theta$) plane is governed by the geodesic equations
\begin{equation}
{\mathscr T}= \pm\int_{r_0}^r\frac{\rd r}{\sqrt{{\cal R}(r)}}=\pm\int_{\theta_0}^{\theta}
              \frac{\rd\theta}{\sqrt{\Theta(\theta)}}
\end{equation}
where ${\mathscr T}$ is the affine parameter and the $\pm$ signs represent the  
increment ($+$) or decrement ($-$) of $r$ and $\theta$ coordinates along the trajectory. The reader is 
recommended to refer to Rauch \& Blandford (1994), Cadez et al. (1998), and Li et al. (2005) 
for a comprehensive description of the analytic solutions
of  Equation (22). Given two constants of motion, $\lambda$ 
and ${\cal Q}$, and the spin parameter $a$, the trajectory of a photon is uniquely determined.
We neglect the photons with trajectories returning to the accretion disk but may encounter the TeV 
photons along their path because of rare probability of occurrence. 

\subsection{The observed emergent spectrum}
The specific flux density observed by a remote observer is expressed as
\begin{equation}
 F_{\nu_{\rm o}}=\int I_{\nu_{\rm o}}\rd\Omega_{\rm o},
\end{equation}
where $I_{\nu_{\rm o}}$ is the specific intensity as a function of frequency $\nu_{\rm o}$ in the observer's 
frame and $\rd\Omega_{\rm o}$ is the element of the solid angle subtended by the image of the accretion disk 
on the observer's sky. In a common way, the image of the disk can be described by two impact parameters 
$\alpha$ and $\beta$ (see Figure~1), which respectively represent the displacement of the image perpendicular 
to the projection of the rotation of the black hole on the sky and the displacement parallel to the projection 
of the axis (Li et al. 2005). Applying the invariant $I_\nu/\nu^3$ along the path of a photon (Rybicki \& Lightman 
1979, p. 146), we have
\begin{equation}
 F_{\nu_{\rm o}}=\int g^{-3}I_{\nu_{\rm e}}\frac{\rd\alpha \rd\beta}{D^2},
\end{equation}
where $\nu_{\rm e}$ is the frequency of the photons in the local rest frame (LRF) of the accretion flows,
$g=\nu_{\rm o}/\nu_{\rm e}$ is the redshift factor of the photons (see Section 3.3), $I_{\nu_{\rm e}}$ is 
the specific intensity of the disk radiation at radius $r_{\rm e}$, and $D$ is the distance of the black hole 
from the observer. Then the observed emergent spectrum is 
\begin{equation}
L_{\nu_{\rm o}}=4\pi D^2F_{\nu_{\rm o}}=4\pi\int g^{-3}I_{\nu_{\rm e}}\rd\alpha \rd\beta.
\end{equation}
For each ($\alpha, \beta$) set, the constants of motion $\lambda$ and $\cal Q$ can be expressed as (Li et al. 2005)
\begin{equation}
\lambda=-\alpha\sin\tobs,~~~~{\cal Q}=\beta^2+(\alpha^2-a^2)\cos^2\Theta_{\rm obs},
\end{equation}
where the viewing angle $\tobs$ is the inclination between the rotation axis of the accretion disk
and the direction to the observer.
Using the ray tracing method we can locate the emission place $r_e$ in the accretion disk for the 
photons which reach the observer's sky at point ($\alpha, \beta$), and, therefore, obtain the radiation intensity 
$I_{\nu_{\rm e}}$ from the GR RIAF model. The observed emergent spectrum
 is obtained by integrating $I_{\nu_{\rm e}}$ over the observer's sky as shown in Equation~(25).

\figurenum{1}
\begin{figure*}[t]
\centering
\includegraphics[angle=-90,width=0.80\textwidth]{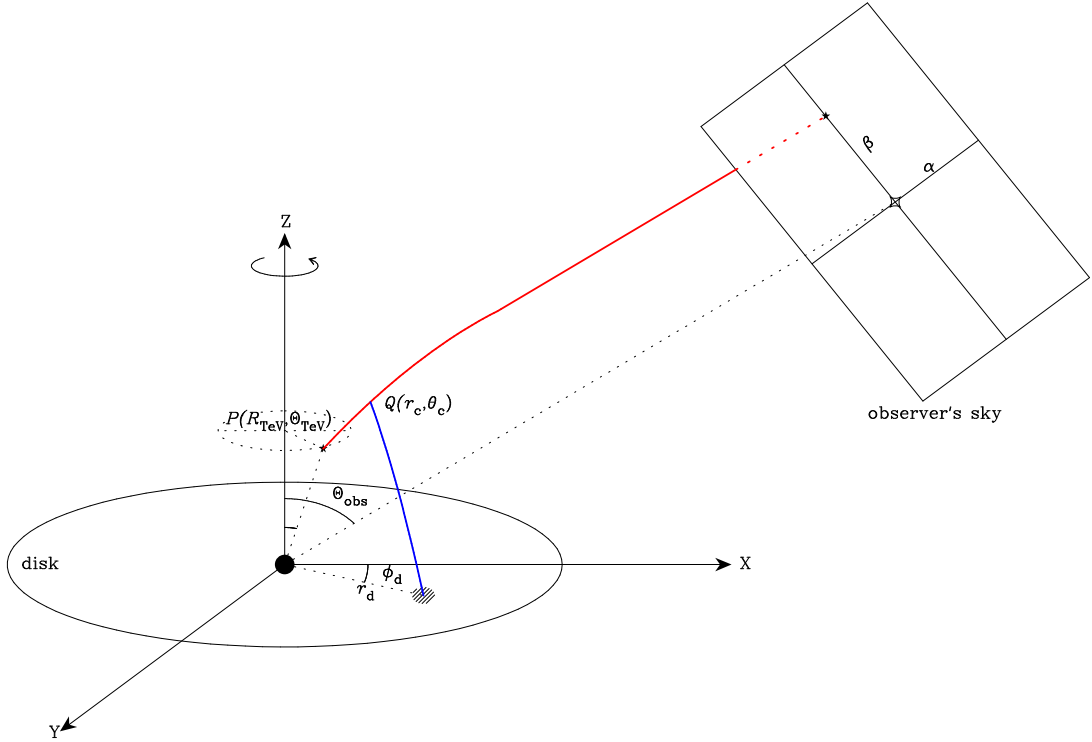}
\caption{\footnotesize Geometric scheme. The point $P(\rtev,\ttev)$ is the location of the 
TeV source and $Q(r_{\rm c},\theta_{\rm c})$ is the location of  the interaction of TeV 
photons with the soft photons from the RIAFs.
}
\label{fig1}
\end{figure*}
\figurenum{2}
\begin{figure}
\centering
\includegraphics[angle=-90,width=0.48\textwidth]{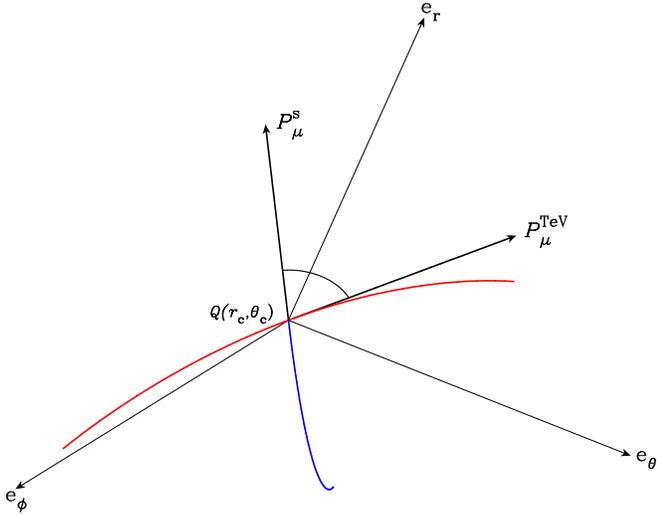}
\caption{\footnotesize Locally nonrotating frame at the interacting point 
$Q(r_{\rm c},\theta_c)$.
}
\label{fig2}
\end{figure} 

\subsection{Optical depth to TeV photons}
Figure~1 shows the geometric scheme for calculations of the optical depth to TeV photons which emanate from the 
position $P(\rtev, \ttev)$. The TeV photons unavoidably interact with the soft photons from the accretion disk 
when they are moving out [e.g., at the interacting point $Q(r_{\rm c}, \theta_{\rm c})$], setting up strong 
constraints on the radiation fields of soft photons. We delineate the interactions in the LNRF (see Figure 2). 
Firstly we divide the solid angle at the interacting point in the LNRF into numerous elements 
$(\Delta \theta_{\rm s}, \Delta\phi_{\rm s})$. 
For each solid angle element at which the soft photons from the disk arrive, we can determine their constants 
of motion $\lambda_{\rm s}$ and ${\cal Q}_{\rm s}$ (see Equation (36)). Secondly we use the ray tracing method 
to trace the soft photons  back to the accretion disk. We hence obtain the number density of soft photons. Lastly 
the optical depth is calculated by integrating all the soft photons from different directions along the trajectory 
of the TeV photons.

Generally speaking, given the emission location of TeV photons ($\rtev$, $\ttev$) and the viewing
angle of an observer ($\tobs$), there is 
a large number of trajectories to the observer at infinity because of the arbitrariness of the $\phi$-component.
Since TeV photons along the shortest path intuitively suffer the least interactions and thus undergo
the minimum optical depth, 
we only consider the optical depth for the shortest path in our calculations.
We select the trajectory with $\alpha=0$ as the shortest path, which represents the case that the TeV photons 
lie at the projection of the rotation axis on the image sky. Keep in mind that genuine emission 
position always has nonzero $\phi$-component, i.e., $\Phi_{\rm TeV}\neq0$, due to the drag of inertia caused 
by the rotation of black hole.

In the LNRF, given the frequency of TeV photons
$\nu_{\infty}^{\rm TeV}$ at infinity, we can determine their corresponding frequency at
the interacting point $Q(r_{\rm c}, \theta_{\rm c})$ by redshift factor defined as
\begin{equation}
g_{_{\rm TeV}}=\frac{\nu_{\infty}^{\rm TeV}}{\nu_{r_c}^{\rm TeV}}
           =\frac{\left.e^\mu_{(t)}P_{\mu}^{\rm TeV}\right|_\infty}
	         {\left.e^\mu_{(t)}P_{\mu}^{\rm TeV}\right|_{r_{\rm c}}}
           =\left.\frac{\Sigma^{1/2}\Delta^{1/2}}{A^{1/2}}\frac{1}{1-\omega\lambda_{\rm TeV}}
            \right|_{r_{\rm c}},
\end{equation}
where the subscript $r_{\rm c}$ or $\infty$ means their values are calculated at position 
($r_{\rm c}, \theta_{\rm c}$) or infinity and similarly hereinafter.
In the same way, at the interacting point the frequency of soft
photons is given by the redshift factor in comparison with their frequency at the emanating place $r_{\rm d}$
in the accretion disk,
\begin{equation}
g_{\rm s}=\frac{\left.e^\mu_{(t)}(\rm LNRF)P_\mu^{\rm s}\right|_{r_{\rm c}}}
               {\left.e^\mu_{(t)}(\rm LRF)P_\mu^{\rm s}\right|_{r_{\rm d}}}
         =\frac{\left.e^{-\nu}(1-\omega\lambda_{\rm s})\right|_{r_c}}
	       {\left.\gamma_r\gamma_\phi e^{-\nu}\left[1-\Omega\lambda_{\rm s}\mp
	       \frac{\displaystyle \beta_r {\cal R}(r)^{1/2}}{\displaystyle  \gamma_\phi A^{1/2}}\right]
	       \right|_{r_{\rm d}}}.
\end{equation}

The directional angles of the TeV and the soft photons can be obtained from the projection of their 
four-momentum onto the spatial directions of the LNRF, respectively. The direction cosines 
($\alpha_{_{\rm TeV}},\beta_{_{\rm TeV}},\gamma_{_{\rm TeV}}$) of the TeV photons in the LNRF are given by
\begin{equation}
\gamma_{_{\rm TeV}} \equiv \cos\theta_{_{\rm TeV}}
                    = \frac{\left.e^\mu_{(r)}P_{\mu}^{\rm TeV}\right|_{r_{\rm c}}}
                          {-\left.e^\mu_{(t)}P_{\mu}^{\rm TeV}\right|_{r_{\rm c}}}
                   =\frac{\pm {\cal R}(r)^{1/2}}{A^{1/2}(1-\omega\lambda_{\rm TeV})},
\end{equation}
\begin{equation}
\beta_{_{\rm TeV}}  \equiv \sin\theta_{_{\rm TeV}}\cos\phi_{_{\rm TeV}}
                   =\frac{\left.e^\mu_{(\theta)}P_{\mu}^{\rm TeV}\right|_{r_{\rm c}}}
                         {-\left.e^\mu_{(t)}P_{\mu}^{\rm TeV}\right|_{r_{\rm c}}}
                   =\frac{\pm\Theta(\theta)^{1/2}\Delta^{1/2}}
                         {A^{1/2}(1-\omega\lambda_{\rm TeV})},
\end{equation}
\begin{equation}
\alpha_{_{\rm TeV}} \equiv \sin\theta_{_{\rm TeV}}\sin\phi_{_{\rm TeV}}
                   =\frac{\left.e^\mu_{(\phi)}P_{\mu}^{\rm TeV}\right|_{r_{\rm c}}}
                         {-\left.e^\mu_{(t)}P_{\mu}^{\rm TeV}\right|_{r_{\rm c}}}
                   =\frac{\lambda_{\rm TeV}}{\sin\theta_{\rm c}}
		    \frac{\Sigma\Delta^{1/2}}{A(1-\omega\lambda_{\rm TeV})},
\end{equation}
and those of the soft photons are given by replacing the superscript (subscript) ``TeV'' with ``s'' 
in the above equations:
\begin{equation}
\gamma_{\rm s} \equiv \cos\theta_{\rm s}=\frac{\left.e^\mu_{(r)}P_{\mu}^{\rm s}\right|_{r_{\rm c}}}
                                       {-\left.e^\mu_{(t)}P_{\mu}^{\rm s}\right|_{r_{\rm c}}},
\end{equation}
\begin{equation}
\beta_{\rm s} \equiv \sin\theta_{\rm s}\cos\phi_{\rm s}
             =\frac{\left.e^\mu_{(\theta)}P_{\mu}^{\rm s}\right|_{r_{\rm c}}}
	           {-\left.e^\mu_{(t)}P_{\mu}^{\rm s}\right|_{r_{\rm c}}},
\end{equation}
\begin{equation}
\alpha_{\rm s} \equiv \sin\theta_{\rm s}\sin\phi_{\rm s}
              =\frac{\left.e^\mu_{(\phi)}P_{\mu}^{\rm s}\right|_{r_{\rm c}}}
	            {-\left.e^\mu_{(t)}P_{\mu}^{\rm s}\right|_{r_{\rm c}}}.
\end{equation}
Note that $\alpha_{\rm s}^2+\beta_{\rm s}^2+\gamma_{\rm s}^2=1$, just giving $\alpha_{\rm s}$ and 
$\beta_{\rm s}$, we can obtain $\lambda_{\rm s}$ and ${\cal Q}_{\rm s}$ from Equations~(33) and (34). 
For simplicity, we define two denotations
\begin{equation}
\mathscr{A}=\frac{\alpha_{\rm s}\sin\theta_{\rm c} A}{\Sigma\Delta^{1/2}},
~~~\mathscr{B}=\frac{\beta_{\rm s}A^{1/2}(1-\omega\lambda_{\rm s})}{\Delta^{1/2}},
\end{equation}
then we can express $\lambda_{\rm s}$ and ${\cal Q}_{\rm s}$ by 
$\alpha_{\rm s}$ and $\beta_{\rm s}$ as
\begin{equation}
\lambda_{\rm s}=\frac{\mathscr{A}}{1+\omega\mathscr{A}},~~~
{\cal Q}_{\rm s}=\mathscr{B}^2-(a\cos\theta_{\rm c})^2+(\lambda_{\rm s}\cot\theta_{\rm c})^2.
\end{equation}
Having obtained $\lambda_{\rm s}$ and ${\cal Q}_{\rm s}$, the ray tracing method is used to determine the emission 
location $r_{\rm d}$ of the soft photons in the accretion disk.

The number density of the soft photons at the interacting point $(r_{\rm c},\theta_{\rm c})$ is
\begin{equation}
n_{\rm ph}(\theta_{\rm s},\phi_{\rm s},\nu_{\rm s},r_{\rm c},\theta_{\rm c})=\frac{I_{\nu_{\rm s}}}{ch\nu_{\rm s}}
=g_s^3\frac{I_{\nu_{\rm d}}}{ch\nu_s},
\end{equation}
where $h$ is the Planck's constant. Here we have applied the invariant $I_\nu/\nu^3$ along the 
path of a photon.  
Finally, the expression for the optical depth of the disk radiation fields to the TeV photons is written as
\begin{equation}
\tau_{_{\rm TeV}}(\rtev,\ttev)=\iiint
                               \sigma_{\gamma\gamma}(\nu_{_{\rm TeV}},\nu_{\rm s},\bar\mu)
			       \frac{I_{\nu_{\rm d}}}{ch\nu_{\rm s}}g_{\rm s}^3\rd\Omega \rd\nu_{\rm s} \rd l,
\end{equation}
where  $\rd l=e^{\nu}\Sigma \rd\mathscr{T}$  is the proper length differential with $\rd \mathscr{ T}$ defined 
to be differential of the affine parameter $\mathscr{ T}$ along the TeV photons' trajectory (see Equation~(22)),
$\rd\Omega=\sin\theta_{\rm s}d\theta_{\rm s}d\phi_{\rm s}$ is the solid angle element in the LNRF, and 
$\bar{\mu}$ is the cosine of angle between the TeV photons and the soft photons. The cross-section of the two 
colliding photons with energy 
$\epsilon_{_{\rm TeV}}=h\nu_{_{\rm TeV}}/m_ec^2$ and $\epsilon_{\rm s}=h\nu_{\rm s}/m_ec^2$
is given by Gould \& Schr\'eder (1967)
\begin{equation}
\sigma_{\gamma\gamma}=\frac{3\sigma_{\rm T}}{16}(1-v^2)\left[(3-v^4)\ln
                      \left(\frac{1+v}{1-v}\right)-2v(2-v^2)\right],
\end{equation}
for pair productions, where $\sigma_{\rm T}$ is the Thompson cross-section and
$v$ is the velocity of electrons and positrons at the center of momentum frame 
in units of $c$ related to $\epsilon_{_{\rm TeV}}$ and $\epsilon_{\rm s}$ through
\begin{equation}
(1-v^2)=\frac{2}{(1-\bar{\mu})\epsilon_s\epsilon_{_{\rm TeV}}}.
\end{equation}
The cross section depends on the energies of the interacting photons and their colliding 
angle. If the two colliding photons run in parallel ($\bar\mu=1$), their interaction disappears.
The 10TeV photons mostly interact with soft photons of $0.05$eV
for header-to-header collisions ($\bar\mu=-1$).  As shown below,
in specific calculations, the energy of soft colliding photons will be modified due to energy 
shift and bending of their trajectories caused by the GR effects.

It should be pointed out that the GR effects on $\tautev$ occur through three factors: (1) changing the global structures 
of RIAFs, as a result, causing dependence of the radiation fields on the spin parameter; (2) modifying the observed 
spectrum from the RIAFs; and (3) bending the trajectories of TeV photons. For different spins of black holes, the 
global structures of RIAFs become significantly different starting $\sim$100 $R_{\rm g}$ inward. 
Once considering the second influence, the observed spectrum is dependent on the viewing angles.
The third influence is nonnegligible unless the trajectories of TeV photons closely approach 
the black hole ($\sim$10 $R_{\rm g}$). These three factors jointly influence the final optical depth to TeV photons.

\section{Numerical Results}
First, we numerically solve the GR RIAF equations for its structure spanning a large space of parameters.
The results are consistent with the numerical solutions in Manmoto (2000). Second,
we calculate the intrinsic spectrum for each case of RIAFs with different accretion rates and spins. 
Third, the ray tracing method is employed to get the spectrum in an observer's frame. 
Last, the optical depth to 10TeV photons is calculated along their path from
$(\rtev, \ttev)$ to the observer's sky at $(\infty, \tobs)$. The constants of motion $\lambda_{\rm TeV}$ and 
$\cal Q_{\rm TeV}$ are obtained according to Equation~(26), where the impact $\alpha$ is set zero corresponding 
to the shortest path and $\beta$ is determined by solving the geodesic equations. 
The properties of structures and emergent spectrum of RIAFs can be found in Manmoto (2000). 
Here we focus on the optical depth of TeV photons. 

Hereinafter we scale accretion rate in the units of the Eddington rate $\dot M=\dot m\dot M_{\rm Edd}$, where
$\dot M_{\rm Edd}=4\pi GM_{\bullet}m_p/\eta\sigma_{\rm T}c$ and $\eta=0.1$ is the radiative efficiency.
We set $M_{\bullet}=3.2\times 10^{9}M_{\odot}$ throughout the paper, and $\alpha_{\rm d}=0.1$,
$\beta_{\rm d}=0.5$, and $\delta=1.0\times10^{-3}$ in this section.

\subsection{spin dependence}
The black hole spins enhance the GR effects on the RIAF structures, the emergent spectrum, and the photon 
trajectories. Detailed calculations show that both the surface density and the temperature of accretion 
flows at fixed radius increase with the spins, consequently, making the radiation more efficient with the spins. 
In light of the drag of the frame, the trajectories of TeV photons will be elongated and the probabilities of 
pair productions are thus enhanced. These two factors together increase $\tautev$ with the spins. However,
the GR influence on the photon trajectories appears within a radius $\sim$10 $R_{g}$ and, therefore, is less important
at relatively large radius.

Figure~3 shows the dependence of $\tautev$ on the black hole spins for given parameters $\dot m=2.5\times10^{-3}$ and
$\tobs=\ttev=30^{\circ}$ at four emission radii of TeV photons. Obviously, $\tautev$ increases with the spins, 
and its dependence on the spins is more tight at smaller radius because the GR 
effects become more significant. We can find that $\tautev\ge 1$ for $a>0.7$ at $R_{\rm TeV}=15 R_{\rm g}$.
On the other hand, the TeV photons from $R_{\rm TeV}=6 R_{\rm g}$ cannot escape from the radiation fields of 
the GR RIAFs for all spin ranges since $\tautev\gg 1$. 
\figurenum{3}
\begin{figure}[t]
\centering
\includegraphics[angle=-90,width=0.4\textwidth]{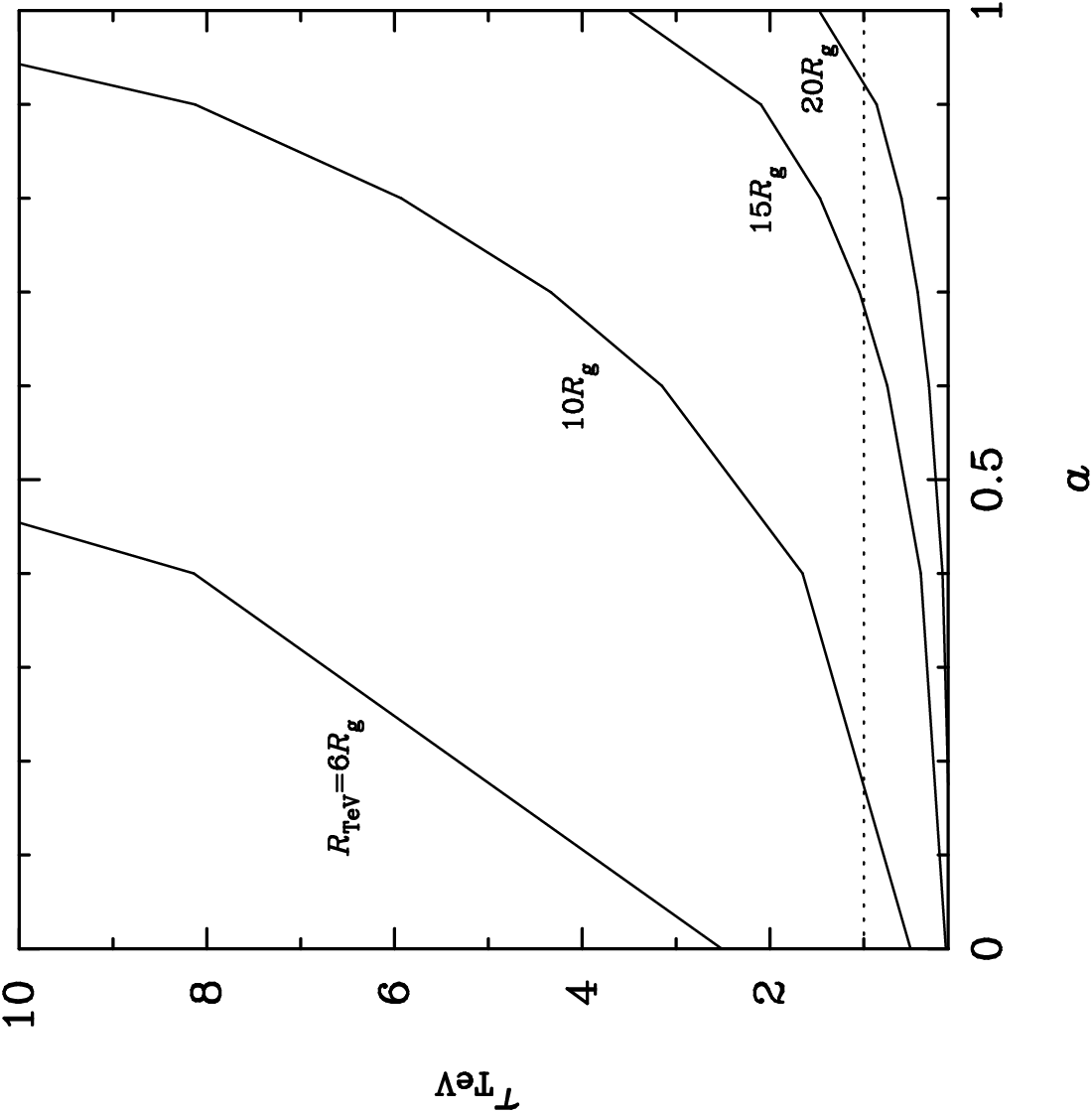}
\caption{Spin dependence of $\tautev$ for parameters $\dot m=2.5\times10^{-3}$ and 
         $\tobs=\ttev=30^{\circ}$ at different $\rtev$.}
\label{fig3}
\end{figure}

\subsection{$R_{\rm TeV}$ dependence}
In Figure 4 we show $\tautev$ as a strong function of $\rtev$, the radius of the location of TeV photons,
by fixing $\dot{m}=2.5\times 10^{-3}$ and $\tobs=\ttev=30^{\circ}$.  The radiation fields from RIAFs become
more and more intensive when $\rtev$ approaches the black hole, leading to a deep dependence of 
$\tautev$ on $\rtev$. It is difficult to get an analytical formulation of the dependence, but Figure 4 shows
the details of the dependence numerically. 
The radiation fields from the RIAFs have different spatial distributions with the spins 
since the GR effects cause their structures distinct. As explained in Section 4.1, the spins enhance the 
radiation fields. If we fix the energy budget released from the RIAFs, the radiation fields
should be more compact for the larger spins, causing a steeper $\tautev$$-$$\rtev$ relation.
The $R_{\rm TeV}$ dependence of $\tautev$ can be applied to estimate the spins.

\subsection{ $\tobs$  and $\ttev$ dependence}
The dependence of $\tautev$ on $\tobs$  and $\ttev$ is caused by the
anisotropy of radiation fields from the disk and the dependence of the cross-section on the colliding
angle in particular. Generally speaking, the number density of soft photons is mostly contributed from the 
inner region of the accretion disk. At larger $\rtev$, the path of these photons that reach the interacting point 
with the TeV photons approximately has the same polar angle with $\ttev$. Since when the two colliding photons 
run in parallel, their interaction disappears. We can see $\tautev$ reaches its minimum at $\tobs=\ttev$, 
whereas at smaller $\rtev$, the TeV photons are embedded in the radiation fields from the inner region of 
the accretion disk and interact with soft photons from all directions and, consequently, it is difficult to give a 
general explanation to the dependence on $\tobs$ and $\ttev$.

We present the $\tautev$-dependence on $\tobs$  and $\ttev$ in Figures 5 and 6, respectively, for
the fixed parameters $\dot{m}=6.3\times10^{-3}$ and $a=0.998$ at different $\rtev$.
From Figure 5 we find that (1) for $\ttev=0^{\circ}$, $\tautev$ monotonously increases with 
$\tobs$, and (2) there is a minimum $\tautev$ for relatively large $R_{\rm TeV}$
at $\tobs=\ttev$. This is confirmed by the case of $\ttev=45^{\circ}$. 
Figure~6 shows that the $\tautev$-dependence on $\ttev$ has the similar properties as that on $\tobs$.

\figurenum{4}
\begin{figure}[t]
\centering
\includegraphics[angle=-90,width=0.40\textwidth]{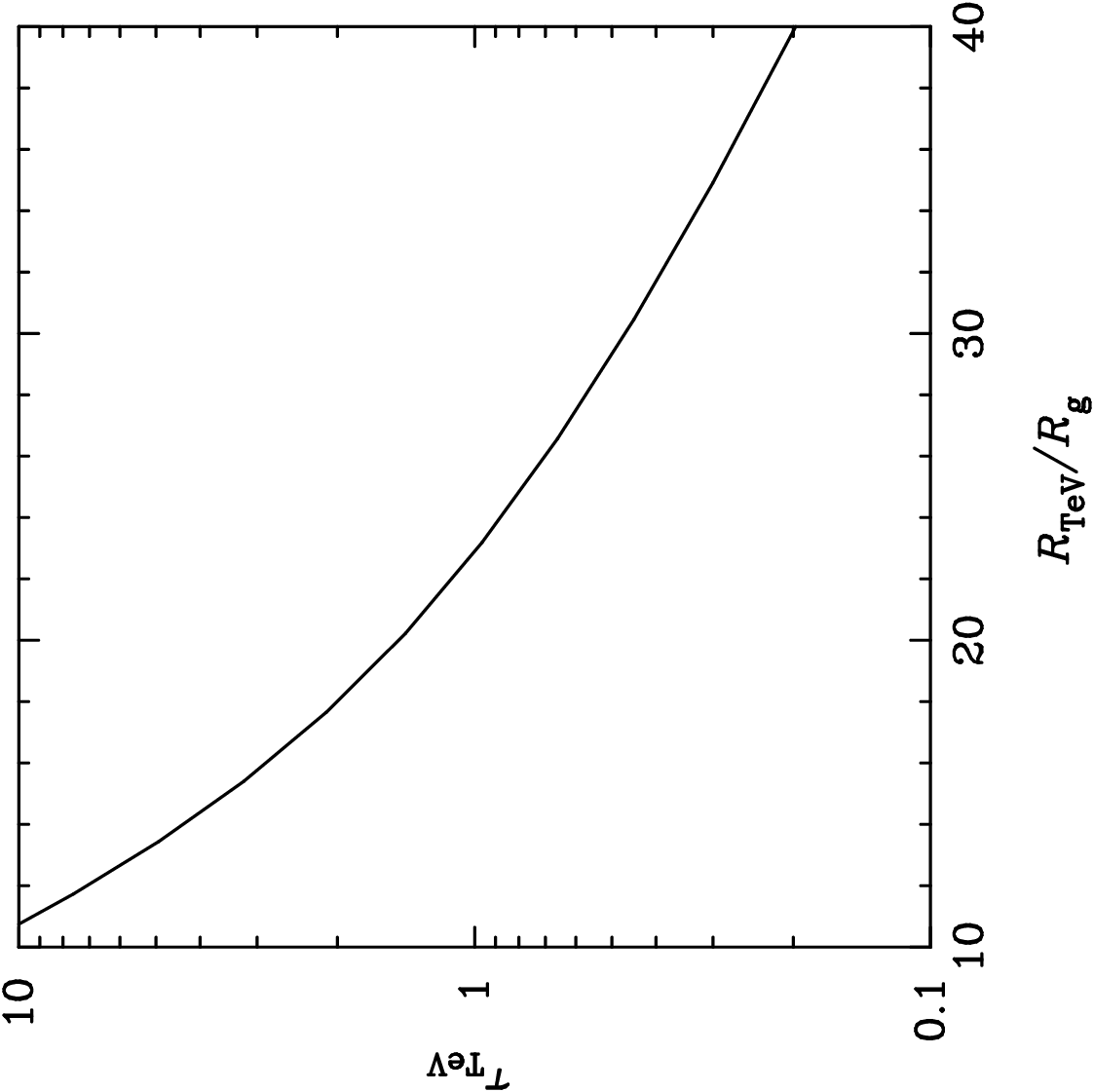}
\caption{$\rtev$ dependence of $\tautev$ for parameters $a=0.998$,
$\dot m=2.5\times10^{-3}$, and $\tobs=\ttev=30^{\circ}$.}
\label{fig4}
\end{figure}

\subsection{ $\dot m$ dependence}
The $\tautev$-dependence on accretion rates is more straightforward in light of the changes of number 
density of soft photons and SED from the RIAFs. 
Figure~7 illustrates how $\tautev$ changes with accretion rates for fixed parameters
$a=0.998$ and $\tobs=\ttev=30^{\circ}$. We find that $\tautev$ is extremely sensitive 
to accretion rates as $\tau_{_{\rm TeV}}\varpropto \dot m^{2.5-5.0}$, and the power index tends to be flat at
larger $\rtev$. This strongly indicates that TeV photons can be violently diluted by minor changes
in accretion rates. In this sense, intensive $\gamma$-ray emission cannot be detected in variable sources. 
This has important implication in observations.

\figurenum{5}
\begin{figure*}[ht]
%\vspace{-5.0em}
\centering
\includegraphics[angle=-90,width=0.70\textwidth]{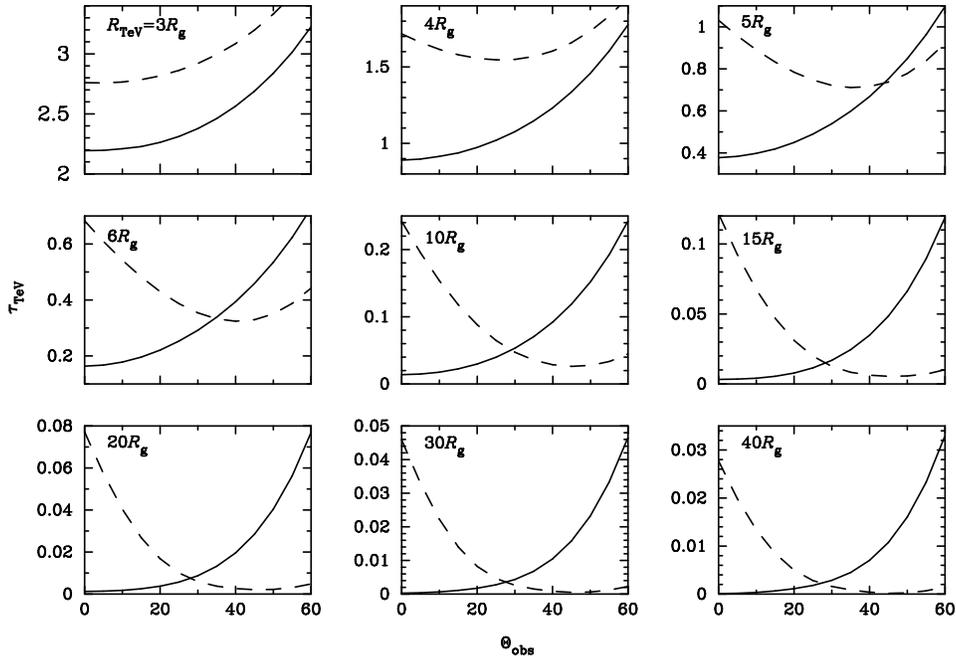}
\caption{\footnotesize $\tobs$ dependence of $\tautev$ for parameters $\dot m=6.3\times10^{-3}$ 
and $a=0.998$. The solid lines are the results for $\ttev=0^{\circ}$ and the dashed lines are the 
ones for $\ttev=45^{\circ}$. 
}
\label{fig5}
\end{figure*}
\figurenum{6}
\begin{figure*}[ht]
%\vspace{-15.0em}
\centering
\includegraphics[angle=-90,width=0.70\textwidth]{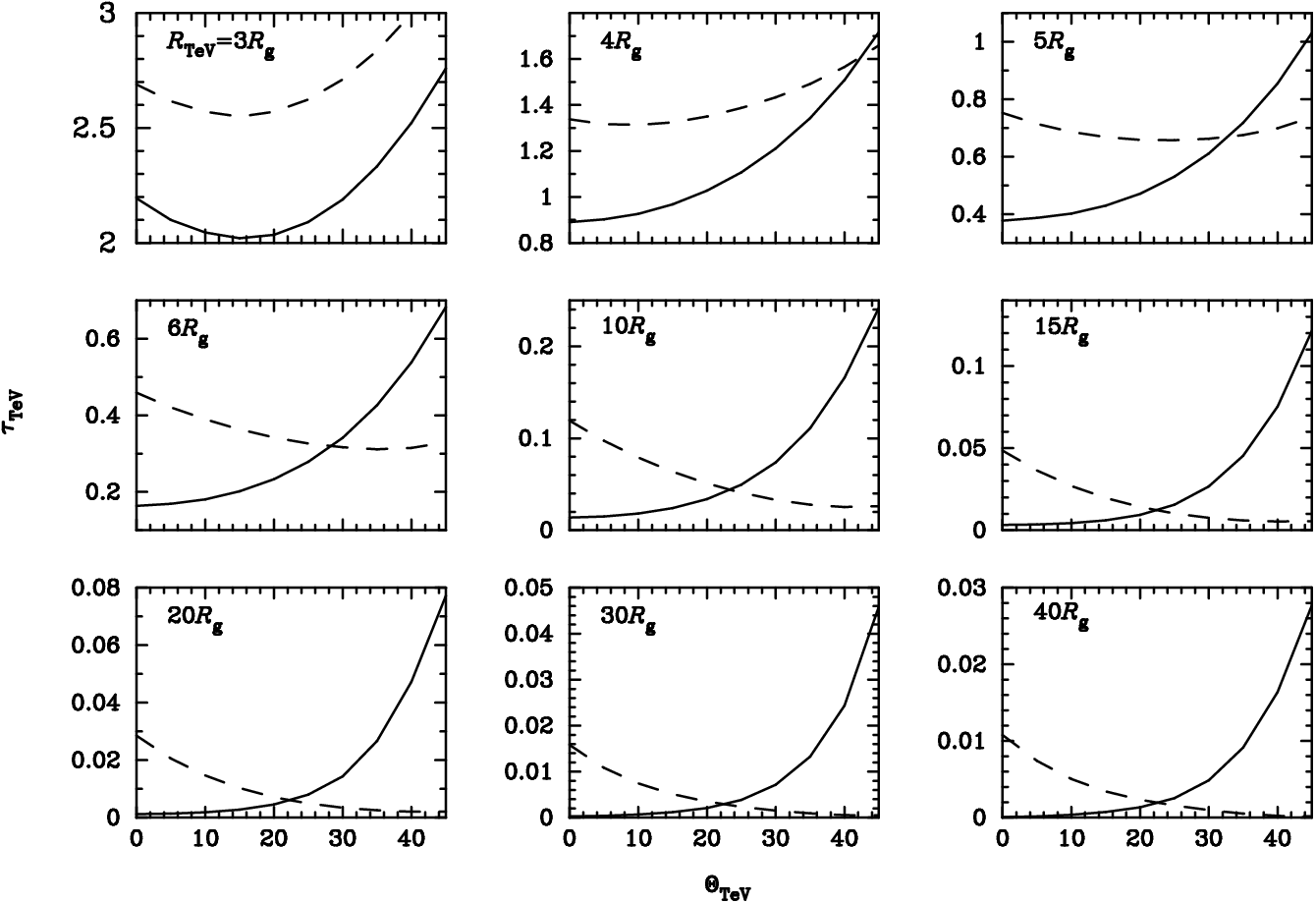}
\caption{\footnotesize $\ttev$ dependence of $\tautev$ for parameters $\dot m=6.3\times10^{-3}$ and $a=0.998$. The solid 
lines are the results for $\tobs=0^{\circ}$ and the dashed lines are the ones for $\tobs=45^{\circ}$.
}
\label{fig6}
\end{figure*}

This strong dependence can be explained in the following ways. Firstly, RIAFs have properties
\begin{equation}
\Sigma_0\sim \dot{m},~~~B\sim \dot{m}^{1/2},~~~
\nu_{\rm syn}\sim \gamma_e^2\dot{m}^{1/2},~~~\tau_{\rm es}\sim \dot{m},
\end{equation}
where $B$ is the magnetic field, $\nu_{\rm syn}$ is peak frequency of synchrotron emission,
$\gamma_e=kT_e/m_ec^2$ is the Lorentz factor of hot thermal electrons,
 and $\tau_{\rm es}$ is the optical depth for Thompson scattering.
The synchrotron emission power {approximates as} $\sim \gamma_e^2\Sigma_0 B^2\sim \gamma_e^2\dot m^2$ and 
the bremsstrahlung emission power $\sim \dot m^2$. In our calculations, we find that $T_e\propto \dot{m}^q$
and hence $\gamma_e\propto \dot{m}^q$, where $q\sim 1/3$.
To understand the $\tautev$-dependence on $\dot{m}$, we have to know the energy of
soft photons interacting with TeV photons. Figure~8 shows the contribution to $\tautev$ for soft photons
with different frequencies from an RIAF model with $\dot{m}=4.0\times10^{-3}$, $a=0.998$, $\tobs=\ttev=30^{\circ}$,
and $R_{\rm TeV}=5R_{\rm g}$. We find that
$\nu_{\rm s}\sim 10^{14.5}$Hz, which corresponds to the first Comptonization peak or the second determined 
by the parameters of the RIAFs.
Since RIAFs are optically thin, most of the synchrotron photons escape and a small fraction 
($\sim \tau_{\rm es}$) are Compton-scattered by the hot electrons, the energy densities
$U_{\rm C_1}\propto \gamma_e^2\dot m^3\propto \dot{m}^{3+2q}$ and 
$U_{\rm C_2}\propto \gamma_e^2\dot m^4\propto \dot{m}^{4+2q}$ for the first and second 
Comptonization, respectively. This determines the sensitivities of $\tautev$-dependence on $\dot{m}$
as
\begin{equation}
\tautev\propto\left\{\begin{array}{l}
U_{\rm C_1}\propto \dot{m}^{3+2q}\sim \dot{m}^{3.7},\\
U_{\rm C_2}\propto \dot{m}^{4+2q}\sim \dot{m}^{4.7}.\end{array}\right.
\end{equation}
For larger $\dot m$, the bremsstrahlung emission
will dominate the synchrotron emission and the Comptonization, even at the lower frequency. This leads 
to a flatter power index $\tau_{_{\rm TeV}}\varpropto \dot m^{2.5}$.
These are nicely consistent with the numerical results as shown in Figure 7.

We would like to point out that the very sensitive dependences of $\tau_{_{\rm TeV}}$
on accretion rate as $\tau_{_{\rm TeV}}\propto \dot m^{2.5-5.0}$ evidently indicate $\tau_{_{\rm TeV}}\gg1$
for the standard accretion disk, immediately drawing a conclusion that the TeV photons cannot escape from
the vicinity of SMBH fueled by the standard accretion disk. This is confirmed by the work of Zhang \& Cheng (1997).

\figurenum{7}
\begin{figure}[t]
\centering
\includegraphics[angle=-90,width=0.40\textwidth]{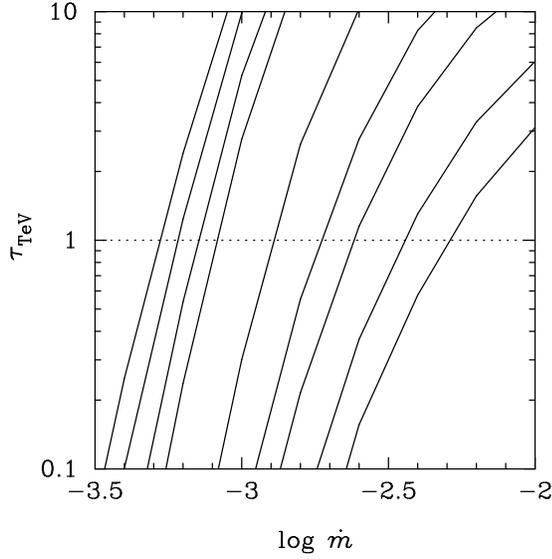}
\caption{$\dot m$ dependence of $\tautev$ for parameters $a=0.998$ and  $\tobs=\ttev=30^{\circ}$. 
From the left to the right, the lines correspond to the results for
$R_{\rm TeV}=3, 4, 5, 6, 10, 15, 20,30,40 R_{\rm g}$, respectively. We can find that 
$\tautev\sim \dot m^{2.5-5.0}$.}
\label{fig7}
\end{figure}
\figurenum{8}
\begin{figure}[t]
\centering
\includegraphics[angle=-90,width=0.40\textwidth]{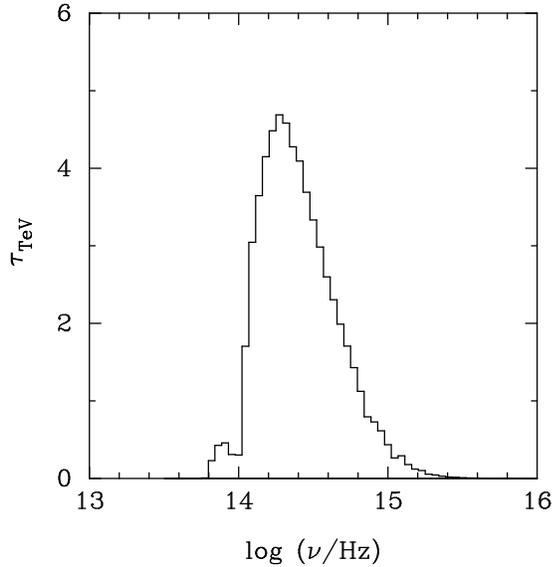}
\caption{Contribution to $\tautev$ from different frequencies of soft photons from the RIAFs with parameters 
$\dot{m}=4.0\times10^{-3}$, $a=0.998$, $\tobs=\ttev=30^{\circ}$ and $\rtev=5R_{\rm g}$.}
\label{fig8}
\end{figure}
\section{Application to M87}
TeV emission with a rapid variability ($\sim 2$ days) has been detected by the HESS in the giant elliptical 
galaxy M87, giving clear evidence for a compact TeV emission region in the immediate vicinity of the black hole 
(Aharonian et al. 2006) and thus providing an opportunity to constrain the spin of the central SMBH.
It is well known that M87 hosts an SMBH with $M_\bullet=(3.2\pm0.9)\times10^{9}M_\odot$ 
(Harms et al. 1994; Macchetto et al. 1997) at a distance of $\sim 16$ Mpc. 
Table 1 summaries the up-to-date observations from radio to X-ray band as shown in Figure 9 and 10. We exclude 
the data with low resolution so as to minimize the contamination from the knots in M87. 
The SED of M87 gives clear evidence for the RIAFs at work in its nucleus (Reynolds et al. 1996; Di Matteo 
et al. 2003).
\begin{deluxetable}{ccccc}
\tabletypesize{\footnotesize}
\tablecolumns{5} 
\tablewidth{0pc} 
\tablecaption{Summary of data for the nucleus of M87.}
\tablehead{ 
\colhead{Frequency}                & \colhead{~$\nu F_{\nu}$~}          &
\colhead{Resolution}               & \colhead{Ref.\tablenotemark{c}} &
\colhead{Obs.~}\\
\colhead{$\nu$(Hz)}                    & \colhead{($10^{-13}$erg s$^{-1}$ cm$^{-2}$)}   &
\colhead{(mas)}               & \colhead{} &
\colhead{}}
\startdata 
\sidehead{\hspace*{0.48\textwidth}Radio}
$5.0\times10^9$      & 0.1\tablenotemark{a}    & 0.7    & 1  & VLBI  \\
$2.2\times10^{10}$   & 0.48\tablenotemark{a}    & 0.15   & 2  & VLBI  \\
$1.0\times10^{11}$   & 8.7\tablenotemark{a}    & 0.1    & 3  & VLBI  \\
\sidehead{\hspace*{0.5\textwidth}IR}
$2.8\times10^{13}$   & $46.4\pm2.5$    &  460   & 7 & Gemini  \\
\sidehead{\hspace*{0.45\textwidth}Optical-UV}
$6.0\times10^{14}$   & $61.7\pm12.3$   & 22     & 6 & FOC, {\em HST}  \\
$7.0\times10^{14}$   & $20.0$\tablenotemark{a}  & 22     & 4,5 & FOS, {\em HST}  \\
$8.1\times10^{14}$   & $37.2\pm7.4$    & 22     & 6 & FOC, {\em HST}  \\
$9.1\times10^{14}$  & $16.0\pm1.6$     & 28.4   & 9 & ACS, {\em HST}  \\
$1.2\times10^{15}$  & $14.0\pm1.4$     & 28.4   & 9 & ACS, {\em HST}  \\
$1.3\times10^{15}$   & $20.4\pm4.0$    & 22     & 6 & FOC, {\em HST}  \\
$1.9\times10^{15}$   & $12.9\pm2.5$    & 22     & 6 & FOC, {\em HST}  \\
$2.0\times10^{15}$   & $16.2\pm3.2$    & 22     & 6 & FOC, {\em HST}  \\
$2.4\times10^{15}$   & $38.9\pm7.8$    & 22     & 6 & FOC, {\em HST}  \\
\sidehead{\hspace*{0.47\textwidth}X-ray}
$2.4\times10^{17}$   & $8.0\pm0.2$\tablenotemark{b}  & 500    & 8 & {\it Chandra}
\enddata 
\tablenotetext{a}{No mention of error-bars.}
\tablenotetext{b}{The X-ray spectral index $\alpha_{_{X}}=-1.23\pm0.04$, defined as $F_{\nu}\propto\nu^{-\alpha_{_X}}$.}
\tablenotetext{c}{References:-
(1) Pauliny-Toth et al. 1981;
(2) Spencer \& Junor 1986;
(3) B\"a\"ath et al. 1992;
(4) Harms et al. 1994;
(5) Reynolds et al. 1996;
(6) Sparks et al. 1996;
(7) Perlman et al. 2001;
(8) Di Matteo et al. 2003;
(9) Maoz et al. 2005.}
\end{deluxetable}
\subsection{Accretion rates of the SMBH}
The radiation fields around the black hole can be quantified from the spectral fitting of M87. 
Before spectral fitting, it is useful to understand  
to what extent the different parameters influence on the structures of the RIAFs and hence their radiation fields. 
Small $\alpha_{\rm d}$ indicates the low efficiency of angular momentum transfer, leading to 
the low radial velocity but high surface density; $\beta_{\rm d}$ represents the fraction of pressure 
contributed by the gas. As to lower $\beta_{\rm d}$, the structures are changed through two ways:
one bt enhancing the magnetic field, and the other by increasing the temperature but reducing the surface 
density; $\dot m$ determines the overall peaks of the spectrum; the parameter 
$\delta$, representing the fraction of the viscous dissipation that heats electrons, plays an important
role in determining the temperature of the electrons and, therefore, in the energy boost of photons after 
Compton-scattering, i.e., $\delta$ mainly determines the frequency location of the Comptonization bumps. 
The value of $\delta$ is still open to question at 
present since some nonthermal mechanisms (e.g., Kolmgorov-like turbulent cascades and collisionless shocks)
are highly unclear, which may significantly change the value of $\delta$ (Narayan \& Yi 1995). 
We treat $\delta$ as a free parameter in our calculations. In the spectral fit, we set $\alpha_{\rm d}$
and $\beta_{\rm d}$ a priori to reduce the freedom. 
\figurenum{9}
\begin{figure*}[t]
\centering
\includegraphics[angle=-90,width=0.85\textwidth]{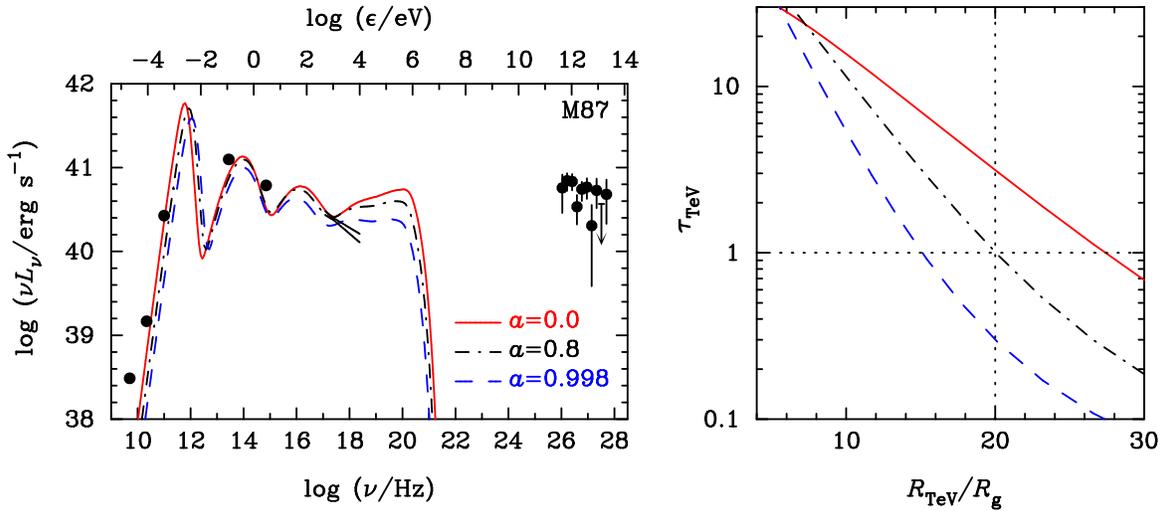}
\caption{\footnotesize { \em Left} panel: The spectral fit of M87 for Case I (see Table 2). 
The data from radio to X-ray band are summarized in Table 1 
and the TeV data are from the HESS observations (Aharonian et al. 2006).
The fit parameters are listed in Table 3. 
{\em Right} panel: The optical depth to 10TeV photons with $\ttev=30^{\circ}$. 
}
\label{fig9}
\end{figure*}
\figurenum{10}
\begin{figure*}[t]
\centering
\includegraphics[angle=-90,width=0.85\textwidth]{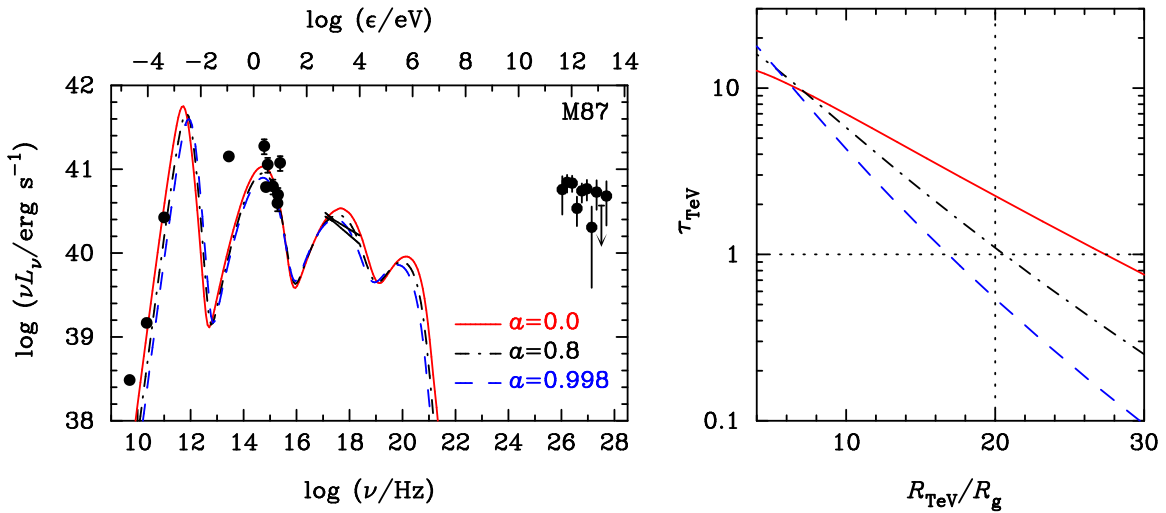}
\caption{\footnotesize Same as in Figure 9 but for Case II (see Table 3).
}
\label{fig10}
\end{figure*}
We use the GR RIAF model to fit the multiwavelength spectrum of M87. 
We firstly fit the observation data used in Paper I (hereafter Case I).
We find that $\alpha_{\rm d}=0.025$ and $\beta_{\rm d}=0.2$ give a good fit to the data. $\dot m$ and 
$\delta$ are adjusted with different spins.
We furthermore take into account the HST observations (Sparks et al. 1996; Maoz et al. 2005) and redo the fit 
(hereafter Case II). 
We set the characteristic value of $\alpha_{\rm d}=0.1$ and $\beta_{\rm d}=0.5$ which means equipartition 
between the gas and magnetized fields.  We present the spectral fit in the {\em left} panels of Figures 9 and 10 
for Case I and Case II, respectively. The fit parameters are listed in Table 3.
The main differences between the two cases are the locations of the multi-Comptonization bumps. In terms of the 
energy dissipation to heat the electrons, the parameters $\alpha_{\rm d}$ and $\delta$ can 
be absorbed into one parameter (see Equation (11)). We note that $\delta\alpha_{\rm d}$ of 
Case II is larger by factor $\sim$4 to that of Case I, which leads to the difference by factor $\sim$4 on
the location of the first Comptonization bump. 

The accretion rates obtained from fitting of the SED in M87 can be examined independently.
Di Matteo et al. (2003) give the Bondi 
accretion rate of the nucleus of M87 using the {\it Chandra} observation as an upper limit of the
accretion rate, $\dot{M}<\dot M_{\rm Bondi}=1.6\times10^{-3}\dot M_{\rm Edd}$. The accretion rate
given by the RIAFs fitting in the current paper is consistent with this upper limit.

We list published literature that gives an estimate of jet power in M87 in Table 2. 
In regard to the inevitable uncertainties of the estimates, the jet power in M87 is in a range of 
$10^{42}$$-$$10^{44}$ erg s$^{-1}$. There are a number of theoretical calculations of jet power driven by BZ 
(Blandford$-$Znajek) or BP (Blandford$-$Payne) mechanisms. For example, Meier (2001) developed a hybrid jet 
model by combining the two processes, in which the jet power is given by
\begin{equation}
\frac{L_{\rm jet}^{\rm Kerr-ADAF}}{L_{\rm acc}}
=0.1\left(\alpha_{-1/2}^{\rm ADAF}\right)^{-1}(0.14f^2+0.74fj+j^2)g^2,
\end{equation}
where $f=\Omega_0/\Omega_{0,\rm NY}$, $j=cJ/GM^2$, $g=B_{\phi,0}/B_{\phi, \rm NY}$, and 
$L_{\rm acc}=0.1\dot{M}c^2$ (see Meier 2001 for details). Taking $f=j=g\sim1$ and $\alpha=0.1$, the jet power 
will be $L_j\approx 0.6L_{\rm acc}$. Indeed, if considering the field-enhancing shear by frame-dragging effects 
(Meier 2001, Nemmen et al. 2007), which are neglected in the self-similar solutions, 
we may have $L_j>L_{\rm acc}$, corresponding to extracting the rotational energy from the black hole 
(the Penrose mechanism). 
The accretion rates obtained in our paper $\dot m\sim10^{-4}$ are able to produce the jet power
$L_j>L_{\rm acc}=3\times 10^{43}$ erg s$^{-1}$, which generally satisfies the jet energy budget 
from observations. 
\begin{deluxetable}{cl} 
\tabletypesize{\footnotesize}
\tablecolumns{2} 
\tablewidth{0pc} 
\tablecaption{The jet power from the published literature}
\tablehead{ 
\colhead{~~~~$L_j$/erg s$^{-1}$~~~~~~~~} & \colhead{~~~~Ref.~~}
} 
\startdata 
$\sim 10^{44}$ & Bicknell \& Begelman (1996)\\
$2\times 10^{43}$ & Reynolds et al. (1996)\\
$\sim 10^{44}$ & Owen et al. (2000) \\
$3\times10^{42}$ & Young et al. (2002) \\
$\sim 10^{44}$ & Stawarz et al. (2006) \\
$5\times10^{43} $ & Bromberg \& Levinson (2008)
\enddata
\end{deluxetable}
\begin{deluxetable}{ccccccc} 
\tabletypesize{\footnotesize}
\tablecolumns{6} 
\tablewidth{0pc} 
\tablecaption{The fit parameters ($\tobs=30^\circ$).}
\tablehead{ 
\colhead{} &
\colhead{~~$a$~~}      & \colhead{~~$\alpha_{\rm d}$~~}   &\colhead{$\beta_{\rm d}$}     & 
\colhead{~~$\delta$~~} & \colhead{~~$\dot m$~~}           & \colhead{$R(\tautev=1)/R_{\rm g}$}
}
\startdata 
        &0.0     & 0.025   & 0.20   & 0.39  &  $2.0\times10^{-4}$  & 28\\
Case I  &0.8     & 0.025   & 0.20   & 0.18  &  $1.8\times10^{-4}$  & 20\\
        &0.998   & 0.025   & 0.20   & 0.09  &  $1.4\times10^{-4}$  & 15\\
\cline{1-7}
        &0.0     & 0.1   & 0.5   & 0.35  &  $1.0\times10^{-4}$  & 28\\
Case II &0.8     & 0.1   & 0.5   & 0.18  &  $9.0\times10^{-5}$  & 20\\
        &0.998   & 0.1   & 0.5   & 0.11  &  $7.5\times10^{-5}$  & 16
\enddata 
\end{deluxetable}
\figurenum{11}
\begin{figure*}[t]
\centering
\includegraphics[angle=-90,width=0.80\textwidth]{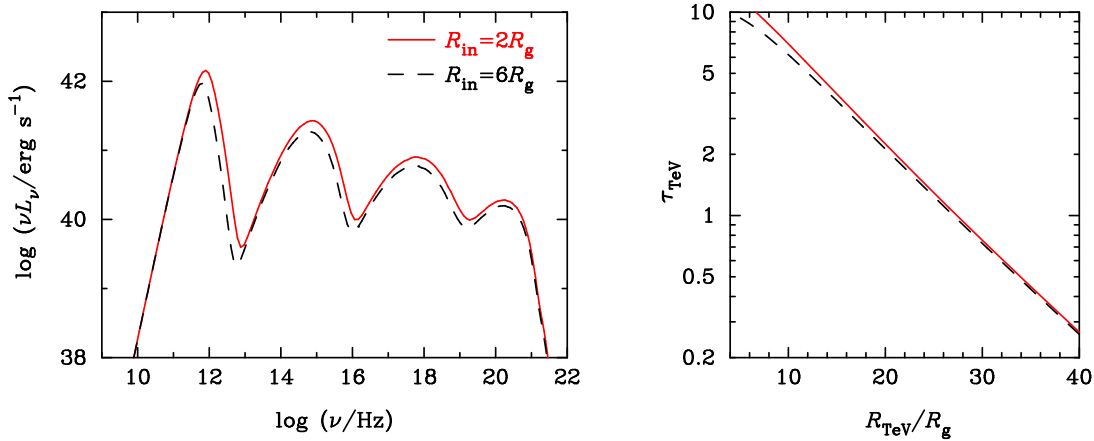}
\caption{\footnotesize Influence of the inner edge of the accretion disk on the spectra and $\tautev$.
The parameters are $a=0.0$, $\alpha_{\rm d}=0.1$, $\beta_{\rm d}=0.5$, $\delta=0.35$, $\dot m=1.0\times10^{-4}$, 
and $\ttev=\tobs=30^{\circ}$. For $a=0.998$, the last stable orbit is equal to the gravitational radius.
}
\label{fig11}
\end{figure*}
\subsection{Spin of the SMBH}
We calculate the optical depth to 10TeV photons and show the results in the {\em right} panel of Figures 9 and 10,
 in which we set $\tobs\sim 30^\circ$ according to the VLBI observation of the jets in M87 (Bicknell 
\& Begelman 1996) since the jets generally align at the axis of the accretion disk. For both cases, the 
resultant $\tautev$$-$$\rtev$ relation 
becomes steeper with the spins. If we define the transparent radius $R_c$ as the radius at which $\tautev=1$, 
we find $R_c\approx15R_{\rm g}$ and $28R_{\rm g}$ for $a=0.998$ and $a=0$ in Case I, respectively, 
and $R_c\approx16R_{\rm g}$ and $28R_{\rm g}$ in Case II. The variability of TeV emission at a timescale 
of $\sim$2days constrains the emission region $\rtev\le20R_{\rm g}$ (Aharonian et al. 2006;  Paper I).
We find that a black hole with a spin of $a_{_{\rm TeV}}=0.8$ for both cases leads to $R_c=20R_{\rm g}$.  
To avoid being optically thick to 10TeV photons, it requires $\tautev\lesssim1$ at $\rtev=20R_{\rm g}$.
In this sense, the spin of SMBH in M87 should be $a \ge 0.8$ 
as shown in Figures 9 and 10.

We have to point out that the quality of the current data does not allow us to ascertain which fit is more 
practical to describe the accretion flows in M87. Future multiwavelength observations with high spatial
resolution will provide stronger constraints on the spectrum and hence on the spin of the 
black hole in M87.

\section{Discussions}
In this paper we fix the horizon of the black hole as the inner edge of the accretion flows. However,
Krolik \& Hawley (2002) show that the inner edge of the accretion flows is dependent on the 
accretion rate and is time-variable. To investigate this effect on the optical depth, we compare the 
optical depth to 10TeV photons for the cases with inner edges set  at the horizon and the last stable 
orbit, respectively. Figure 11 shows the results with parameters $a=0$ (the horizon is $2R_{\rm g}$ and the 
last stable orbit is $6R_{\rm g}$), $\alpha_{\rm d}=0.1$, $\beta_{\rm d}=0.5$, $\delta=0.35$, 
$\dot m=1.0\times10^{-4}$, and $\ttev=\tobs=30^\circ$.
We can clearly see from Figure 11 that the inner edge has little influence on the optical depth
of the TeV photons.
We presume that the location of TeV origination
should be close to the horizon of the black hole.  This provides strong constraints
on the radiation fields near the black hole horizon. With the detailed calculations presented in this paper, there 
are still some aspects to be included in the future work.

(1) {\em The size of the TeV source}. We simply assume a point-like TeV emission source and neglect its size.
Once considering spatial distribution of the source, $\tautev$ should depend on the location of
TeV photons since they have different trajectories.
We here only focus on the minimum optical depth; however, the results in the 
present paper are viable for the current instruments of TeV detection. 
Future work on the size effects could produce interesting results on variabilities of 
TeV photons, including the profile of light curves and the time lag between the different TeV photons. 
This may provide a mapping of the TeV  source so that the radiation mechanism will be finally discovered 
(B\"ottcher \& Dermer 1995; Levinson 2000; Neronov \& Aharonian 2007; Rieger \& Aharonian 2008).

(2) {\em The motion of the TeV source}. 
For TeV photons with specified trajectory, their optical depth is independent of the properties of the 
TeV source. However, the beaming effects caused by relativistic motion of the TeV source will help to avoid 
pair-production absorption of TeV photons, in regard to the beaming of the intensity of TeV 
photons along the direction of the relativistic motion, and the blueshift of the energy of TeV photons 
by comparing with that in the source frame. This will modify the observed spectra of TeV emission.
Interestingly, the present procedure modified by including beamed TeV photons from relativistic jets can be applied
to blazars, which are being powered by the standard accretion disks 
(in flat spectrum radio quasars) and the RIAFs (in high-frequency peaked BL Lacs, such as Mkn 421 and Mkn 501) 
(Wang et al. 2003).

(3) {\em The vertical structure of the RIAFs}. An exponential profile of the density has been adopted in the 
vertical direction of the RIAFs, which means that most soft photons are emanating from the midplane.
Future three-dimensional simulations of the accretion flows will give a more realistic description of 
the RIAF structures. In this sense, the optical depth of the TeV photons should be calculated in a more 
sophisticated way. 

(4) {\em The time-dependent trajectories of TeV photons}. Future detectors with large area will receive
more TeV photons and be able to give more details of spectral variabilities. This needs a consideration on 
the time-dependent trajectories of TeV photons in light of $\tautev$ (time-dependent pair productions when
$\tautev>1$), for example, the periodic rotation of the TeV source around the central SMBH.
This will definitely provide much more information about the innermost region near the 
black hole's horizon.

In other words, further detailed theoretical work is needed for stronger limits on SMBH spins from 
future observations. The current calculations based on the GR RIAF model provide valuable constraints on
the target fields of TeV photons and therefore on the spins.

\section{Summary}
The optical depth to energetic TeV photons, which are immersed in the radiation fields from radiatively inefficient 
accretion flows, has been calculated in detail by including all the GR effects. We investigate the dependence 
of optical depth on the spins ($a$), accretion rates ($\dot m$), viewing angles ($\tobs$), and location 
of the TeV photons ($\rtev, \ttev$). We find that the optical depth is more sensitive to $\rtev$ than to
$\tobs$ and $\ttev$. One of the most interesting results is that $\tautev$ strongly depends on the accretion 
rates as $\tautev\propto\dot m^{2.5-5.0}$. 

Applying the dependence of optical depth on $\rtev$ to constrain the spin parameter in M87, wherein the RIAFs
are expected to be at work,
we find that the observed TeV photons detected by HESS can escape from the radiation fields from the RIAFs with 
spin $a\gtrsim0.8$. Future observations of the phase II HESS with
threshold energy one order higher than that of the phase I may discover more M87-like objects (e.g., low-luminosity
AGNs) with TeV emission. Hopefully, we then will have a sample for the statistic sample of spins of SMBHs.

\acknowledgements{We thank F. Yuan for his kind help in calculations of emergent spectrum from the RIAFs.
We appreciate the stimulating discussions among the members of IHEP AGN group. The 
research is supported by NSFC and CAS via NSFC-10325313, 10821061, 10521001, 10673010, 10573016, 
KJCX2-YW-T03, the 973 project (No. 2009CB824800) and Program for New Century Excellent Talents
in University, respectively. The codes used in this paper are available
for readers interested in calculations via email to {\em liyanrong@mail.ihep.ac.cn}.}

\appendix
\section{ Preliminaries of GR notations}
We adopt geometrical units ($G=c=1$) throughout the Appendix, where $G$ is the gravitational constant 
and $c$ is the light speed. We use the Kerr metric in Boyer$-$Lindquist coordinates ($t, r, \phi, \theta$)
\begin{equation}
\rd s^2=-e^{2\nu}\rd t^2+e^{2\psi}(\rd\phi-\omega \rd t)^2+e^{2\mu_1}\rd r^2+e^{2\mu_2}\rd\theta^2,
\end{equation}
with
\begin{equation}
e^{2\nu}=\frac{\Sigma \Delta}{A},~~~e^{2\psi}=\frac{\sin^2\theta A}{\Sigma},
\end{equation}
\begin{equation}
e^{2\mu_1}=\frac{\Sigma}{\Delta},~~~e^{2\mu_2}=\Sigma,~~~\omega=\frac{2M_{\bullet}ar}{A},
\end{equation}
and
\begin{equation}
\Delta=r^2-2M_{\bullet}r+a^2,~~~\Sigma=r^2+a^2\cos^2\theta,~~~A=(r^2+a^2)^2-a^2\Delta\sin^2\theta,
\end{equation}
where $a=J/M_{\bullet}$ is the specific angular momentum and $M_{\bullet}$ 
is the black hole mass. The event horizon lies at
\begin{equation}
 r_{\rm h}=M_{\bullet}+(M_{\bullet}-a^2)^{1/2}.
\end{equation}
The angular frequencies of the corotating (+) and counterrotating ($-$) Keplerian motions are
\begin{equation}
\Omega_{\rm K}^{\pm}=\pm\frac{M_{\bullet}^{1/2}}{r^{3/2}\mp aM_{\bullet}^{1/2}}.
\end{equation}

To describe the motions of the accretion flows or photons in Kerr metric, we employ three  
reference frames (Gammie \& Popham 1998). 
The first is the LNRF, an orthonormal tetrad basis carried by 
observers who live at constant $r$ and $\theta$, but at $\phi=\omega t~+~ $constant; 
the second is the CRF, whose coordinate angular velocity is 
$\Omega$, here $\Omega$ is the angular velocity of the accretion flows. The last is 
the LRF of the accretion flows. The basis vectors for the LNRF are
\begin{equation}
e^\mu_{(t)}=e^{-\nu}(1,0,0,\omega),
\end{equation}
\begin{equation}
e^{\mu}_{(r)}=e^{-\mu_1}(0,1,0,0),
\end{equation}
\begin{equation}
e^{\mu}_{(\theta)}=e^{-\mu_2}(0,0,1,0),
\end{equation}
\begin{equation}
e^{\mu}_{(\phi)}=e^{-\psi}(0,0,0,1),
\end{equation}
and for the LRF are
\begin{equation}
e^{\mu}_{(t)}=(\gamma_r\gamma_\phi e^{-\nu},~~\gamma_r\beta_re^{-\mu_1},~~0,
               ~~\gamma_r\gamma_\phi e^{-\nu}\Omega),
\end{equation}
\begin{equation}
e^{\mu}_{(r)}=(\gamma_r\gamma_\phi \beta_r e^{-\nu},~~\gamma_r e^{-\mu_1},~~0,
               ~~\gamma_r\gamma_\phi\beta_r e^{-\nu}\Omega),
\end{equation}
\begin{equation}
e^{\mu}_{(\theta)}=(0,~~0,~~e^{-\mu_2},~~0),
\end{equation}
\begin{equation}
e^{\mu}_{(\phi)}=\left[\gamma_\phi\beta_\phi e^{-\nu},~~0,~~0,
                 ~~\gamma_\phi(\beta_\phi\omega e^{-\nu}+e^{-\psi})\right],
\end{equation}
where $\beta_r$ is the radial velocity of the accretion flows in the CRF 
with $\gamma_r=(1-\beta_r^2)^{-1/2}$,
$\beta_\phi=e^{\psi-\nu}(\Omega-\omega)$ is the  physical azimuthal velocity of the 
CRF with respect to the LNRF with $\gamma_\phi=(1-\beta_\phi^2)^{-1/2}$.


\begin{thebibliography}{}
\bibitem[Abramowicz et al. (1996)]{abra96}
Abramowicz, M. A., Chen, X., Granath, M. \& Lasota, J. P. 1996, \apj, 471, 762
\bibitem[Abramowicz et al. (1997)]{abra96}
Abramowicz, M. A., Lanza, A., Percival, M. J. 1997, \apj, 479, 179
\bibitem[Aharonian et al. (2006)]{ahar06}
Aharonian, A. et al. 2006, Science, 314, 1424
\bibitem[B\"a\"ath et al. (1992)]{baat92}
B\"a\"ath, B. et al. 1992, \aap, 257, 31
\bibitem[Bardeen et al. (1972)]{bard72}
Bardeen, J. M., Press, W. H. \& Teukolsky, S. A. 1972, \apj, 178, 347
\bibitem[Bicknell \& Begelman (1996)]{bick96}
Bicknell, G. V. \& Begelman, M. C. 1996, \apj, 467, 597
\bibitem[Blandford \& Levinson (1995)]{blan95}
Blandford, R. D. \& Levinson, A. 1995, \apj, 441, 79
\bibitem[B\"ottcher \& Dermer (1995)]{bott95}
B\"ottcher, M. \& Dermer, C. D. 1995, \aap, 302, 37
\bibitem[Brenneman \& Reynolds (2006)]{bren06}
Bromberg, O. \& Levinson, A. 2008, arXiv:0810.0562
\bibitem[Cadez et al. (1998)]{cade98}
Cadez, A., Fanton, C., \& Calvani, M. 1998, New Astron., 3, 647
\bibitem[Coppi \& Blandford (1990)]{copp90}
Coppi, P. S. \& Blandford, R. D. 1990, \mnras, 245, 453
\bibitem[Di Matteo et al. (2003)]{dima03}
Di Matteo, T., Allen, S. W., Fabian, A. S., Wilson, A. S. \& Yong, A. J. 2003, \apj, 582, 133
\bibitem[Doeleman et al. (2008)]{doel08}
Doeleman, S. et al., 2008, \nat, 455, 78
\bibitem[Elvis et al. (2002)]{elvi02}
Elvis, M., Risaliti, G. \& Zamorani, G., 2002, ApJ, 565, L75
\bibitem[Fabian \& Iwasawa (1999)]{Fabi99}
Fabian, A. \& Iwasawa, K. 1999, \mnras, 303, L34
\bibitem[Fabian et al. (2002)]{fabi02}
Fabian, A. et al. 2002, \mnras, 335, L1
\bibitem[Gammie \& Popham (1998)]{gamm98}
Gammie, C. F. \& Popham, R. 1998, \apj, 498, 313
\bibitem[Gould et al. (1967)]{goul67}
Gould, R. J. \& Schr\'eder, G. P. 1967, Phy. Rev., 155, 1404
\bibitem[Harms et al. (1994)]{harm94}
Harms, R. J. et al. 1994, \apj, 435, L35
\bibitem[Ho (2008)]{ho08}
Ho, L. C. 2008, \araa, 46, 475
\bibitem[Kaspi et al. (2000)]{kaps00}
Kaspi, S., Smith, P. S., Netzer, H., Maoz, D., Jannuzi, B. \& Giveon, U. 2000, \apj, 533, 631
\bibitem[Koemendy \& Gebhardt (2001)]{korm01}
Kormendy, J. \& Gebhardt, K. 2001, in AIP Conf. Ser. 586, 20th Texas Symposium on Relativistic 
            Astrophysics, ed.  J. C. Wheeler \& H. Martel (New York: AIP), 363
\bibitem[Krolik & Hawley (2002)]{krol02}
Krolik, J.H. \& Hawley, J.F. 2002, \apj, 573, 754
\bibitem[Levinson (2000)]{levi00}
Levinson, A. 2000, \prl, 85, 912
\bibitem[Levinson (2006)]{levi06}
Levinson, A. 2006, Int. J. Mod. Phys. A. 21, 6015 
\bibitem[Li et al. (2005)]{li05}
Li, L.-X., Zimmerman, E. R., Narayan, R. \& McClintock, J. E. 2005, \apjs, 157, 335
\bibitem[Macchetto et al. (1997)]{macc97}
Macchetto, F., Marconi, A., Axon, D. J., Capetti, A., Sparks, W. \& Crane, P. 1997, \apj, 489, 579
\bibitem[Manmoto (2000)]{manm00}
Manmoto, T. 2000, \apj, 534, 734
\bibitem[Manmoto et al. (1997)]{manm97}
Manmoto, T., Mineshige, S. \& Kusunose, M. 1997, \apj, 489, 791
\bibitem[Maoz et al. (2005)]{maoz05}
Maoz, D., Nagar, N. M., Falcke, H. \& Wilson, A. S. 2005, \apj, 625, 699
\bibitem[Meier (2001) ]{meie01}
Meier, D. L. 2001, ApJ, 548, L9
\bibitem[Miller (2007)]{mill07}
Miller, J.~M.\ 2007, \araa, 45, 441
\bibitem[Narayan \& Yi (1995)]{nara95}
Narayan, R. \& Yi, I. 1995, \apj, 452, 710
\bibitem[Nemmen et al. (2007)]{nemm07}
Nemmen, R. S., Bower, R. G., Babul, A., Storchi-Bergmann, T. 2007, \mnras, 377, 1652
\bibitem[Neronov \& Aharonian (2007)]{nero07}
Neronov, A. \& Aharonian, F. A. 2007, \apj, 671, 85
\bibitem[Owen et al (2000)]{owen00}
Owen, F. N., Eilek, J. A. \& Kassim, N. E. 2000, \apj, 543, 611
\bibitem[Pauliny-Toth et al . (1981)]{paul81}
Pauliny-Toth, I. I. K., Preuss, E., Witzel, A., Graham, D., Kellerman, K. I. \& R\"onn\"ag, B.
            1981, \aj, 86, 371
\bibitem[Perlman et al. (2001)]{perl01}
Perlman, E. S., Sparks, W. B., Radomski, J., Packham, C., Fisher, R. S., Pi\~na, R. \& Biretta, J. A.
            2001, \apj, 561, L51
\bibitem[Rauch \& Blandford (1994)]{rauc94}
Rauch, K.~P. \& {Blandford}, R.~D. 1994, \apj, 421, 46
\bibitem[Reynolds et al. (1996)]{reyn96}
Reynolds, C. S., Fabian, A. C., Celotti, A. \& Rees, M. J. 1996, \mnras, 283, 873
\bibitem[Rieger \& Agaronian (2008)]{rieg08}
Rieger, F. M. \& Aharonian, F. A. 2008, \aap, 479, L5 
\bibitem[Rybicki \& Lightman (1979)]{rybi79}
Rybicki, G. B. \& Lightman, A. P. 1979, Radiative Processes in Astrophysics 
           (New York: John Wiley \& Sons)
\bibitem[Shen et al. (2005)]{shen05}
Shen, Z.-Q., Lo, K.Y., Liang, M.-C, Ho, P. P. \& Zhao, J. H. 2005, \nat, 438, 62
\bibitem[Sparks et al. (1996)]{spar96}
Sparks, W. B., Biretta, J. A. \& Macchetto, F. 1996, \apj, 473, 254
\bibitem[Spencer \& Junor (1986)]
{spen86}Spencer, R. E. \& Junor, W. A 1986, \nat, 321, 753
\bibitem[Stawarz et al. (2006)]{staw06}
Stawarz, L., Aharonian, F., Kataoka, J., Ostrowski, M., Siemiginowska, A. \& Sikora, M. 2006, 
            \mnras, 370, 981
\bibitem[Wang et al. (2006)]{wang06}
Wang, J.-M., Chen Y.-M., Ho, L. C. \& McLure, R. 2006, \apj, 642, L111
\bibitem[Wang et al. (2008)]{wang08}
Wang, J.-M., Li Y.-R., Wang, J.-C. \& Zhang, S. 2008, \apj, 676, L109 (Paper I)
\bibitem[Wang et al. (2003)]{wang03}
Wang, J.-M., Staubert, R. \& Ho, L. C. 2003, \aap, 409, 887 
\bibitem[Young et al. (2002)]{youn02}
Young, A. J., Wilson, A. S. \& Mundell, C. G. 2002, \apj, 579, 560
\bibitem[Yu \& Tremaine (2002)]{yu02}
Yu, Q. \& Tremaine, S., 2002, \mnras, 335, 965
\bibitem[Yuan et al. (2009)]{yuan09} 
Yuan, Y.-F., Cao, X., Huang, L., \& Shen, Z.-Q. 2009, ApJ in press (arXiv:0904.4090)
\bibitem[Zhang \& Cheng (1997)]{zhan97}
Zhang, L. \& Cheng, K. S. 1997, \apj, 475, 534
\end{thebibliography}
\end{document}